\documentclass[final,3p,times,twocolumn]{elsarticle}
\usepackage{amssymb}
\usepackage{amsmath}
\usepackage{graphicx}
\usepackage{dcolumn}
\usepackage{bm}

\input{tcilatex}
\begin{document}

\begin{frontmatter}
\title{SU(1,1) interferometry with parity measurement}

\author[]{Shuai Wang\corref{cor1}}
\ead{wshslxy@jsut.edu.cn}
\author[]{Jiandong Zhang}

\cortext[cor1]{Corresponding author}

\address{School of Mathematics and Physics, Jiangsu University of Technology, Changzhou 213001, P.R. China}

\begin{abstract}
We present a new operator method in the Heisenberg representation to obtain
the signal of parity measurement within a lossless SU(1,1) interferometer.
Based on this method, it is convenient to derive the parity signal directly
in terms of input states, including general Gaussian or non-Gaussian state.
As applications, we revisit the signal of parity measurement within an
SU(1,1) interferometer when a coherent or thermal state and a squeezed
vacuum state are considered as input states. In addition, we also obtain the
parity signal of a Fock state when it passes through an SU(1,1)
interferometer, which is also a new result. Therefore, the operator method
proposed in this work may bring convenience to the study of quantum
metrology, particularly the phase estimation based on an SU(1,1)
interferometer.
\end{abstract}


\end{frontmatter}

\section{Introduction}

Over the past decades, optical interferometers have been widely used to
estimate very small phase shifts in both theoretical and experimental
studies on quantum metrology. For a Mach-Zehnder interferometer (MZI) with
nonclassical input states \cite{1,2,3,4,5,6,7,8,9,10}, the sensitivity of
the phase estimation can surpass the shot-noise limit (SNL), $\Delta \phi =1/%
\sqrt{\bar{n}}$\cite{1}, even approach the so-called Heisenberg limit (HL) $%
\Delta \phi =1/\bar{n}$ \cite{11,12}, where $\bar{n}$ is the mean number of
photons inside the interferometer. An MZI is also called an SU(2)
interferometer, as Yurke et al. in 1986 showed that the group SU(2) can
naturally describe an MZI \cite{13}. In addition, in the same paper, the
authors first proposed another type of interferometer characterized by the
group SU(1,1), as opposed to the SU(2) interferometer, where the 50:50 beam
splitters in a traditional MZI are replaced by the nonlinear beam splitters,
such as optical parameter amplifiers (OPA) or four-wave mixing. It can be
shown that an SU(1,1) interferometer, under ideal conditions and in the
large $\bar{n}$ limit, can achieve the HL even if inputs are both vacua,
thus holding out the promise of substantial improvement over the SNL.

For various optical interferometers proposed to improve the phase
sensitivity, they mainly differ in the light they use and, as a consequence,
the measurement scheme that is required for extracting the phase information 
\cite{14}. In general, the phase sensitivity within these settings crucially
depends on the input states. By making adjustments to the measurement
scheme, the Cram\'{e}r-Rao bound may be approached, which is an ultimate
limit on the phase sensitivity given by quantum Fisher information \cite{15}
and only depends on the input states. Besides the intensity measurement and
the balanced homodyne measurement, it has been also shown that the parity
measurement \cite{16} can also reach the Cram\'{e}r-Rao bound for an MZI
with a wide range of input states \cite{17}. Actually, the parity
measurement is to perform photon number parity (the evenness or oddness)
measurements on one of the output modes of the interferometer.
Mathematically, the parity measurement is described by a simple, single-mode
operator,$\,$%
\begin{equation}
\hat{\Pi}=(-1)^{\hat{N}}  \label{a}
\end{equation}%
where $\hat{N}$ is a photon number operator. According to the results in
Ref. \cite{18}, parity measurement satisfies $\left\langle \hat{\Pi}%
\right\rangle =\pi W\left( 0,0\right) $, i.e., the expectation value of the
parity operator can be obtained by calculating the Wigner function of the
output state. Furthermore, based on the fact that the Wigner functions of
unknown quantum states are typically reconstructed after optical quantum
state tomography \cite{19}, Plick et al. \cite{20} in 2010 by the homodyne
measurement presented a method for directly obtaining the parity of a
Gaussian state of light without photon-number-resolving measurement.

In recent years, with the help of the transformation of phase space $W_{%
\text{out}}\left( \alpha ,\beta \right) =W_{\text{in}}\left( \tilde{\alpha},%
\tilde{\beta}\right) $, many studies have been done to investigate the phase
sensitivity with Gaussian or non-Gaussian states considered as the input
states of an MZI interferometer \cite{21,22,23,24,25,26}. On the other hand,
the phase sensitivity for an SU(1,1) interferometer with some Gaussian input
states has also been investigated by the same method \cite{27,28,29}.
However, it is difficult to obtain the parity when a non-Gaussian state
passes through an SU(1,1) by the transformation of phase space \cite{20,21}
or the previous traditional operator method in the in the Schr\"{o}dinger
representation \cite{16}. Different from the previous work, in this paper we
present a new operator method in the Heisenberg representation to obtain the
parity of the output modes of an SU(1,1) interferometer directly in terms of
the input state. Our method is relatively simpler for general input states
including Gaussian and non-Gaussian states.

The structure of the present paper is as follows: In Sec. II. A, we first
introduce the normal ordering form of the unitary operator, which describes
a whole lossless SU(1,1) interferometer by the techniques of integration
within an ordered product of operators (IWOP). By the similar way, we
further introduce a Hermitian operator $\hat{\mu}\left( \xi ,\phi \right) $
which can be completely described the whole operation of the parity
measurement combined with an SU(1,1) interferometer in Sec. II. B. As a
consequence, the signal of the parity measurement within the interferometer
can be expressed by $\left\langle \hat{\Pi}\left( \phi \right) \right\rangle
=$Tr$\left[ \rho _{\text{in}}\hat{\mu}\left( \xi ,\phi \right) \right] $
where $\rho _{\text{in}}$ is an input state. In Sec. III, in order to prove
the superiority of our method, we directly obtain the signals of the parity
measurement within an SU(1,1) interferometer for some Gaussian or
non-Gaussian states.

\section{Equivalent Hermitian operator of parity measurement combined with
an SU(1,1) interferometer}

It is known that, for an SU(1,1) optical interferometer, it is like an MZI
with the beam splitters replaced by two OPAs. Different from the previous
work, here we consider the concrete measurement method (for example, the
parity measurement) and an SU(1,1) interferometer as a whole operation which
can be represented by a Hermitian operator as shown in Fig. 1. In this way,
one can obtain the signal of the parity measurement within an SU(1,1)
interferometer directly in terms of the input state. 
\begin{figure}[tbph]
\centering
\includegraphics[width=8cm]{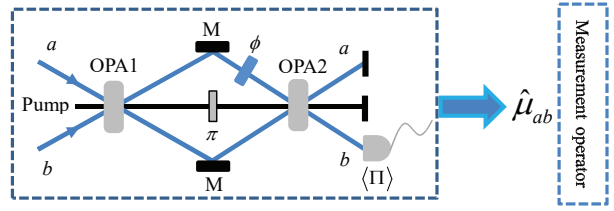}
\caption{(color online) Sketch of parity measurement of a two-mode quantum
state passing through an SU (1,1) interferometer.}
\end{figure}

\subsection{A. Normal ordering form of the unitary operator corresponding to
a lossless SU(1,1) interferometer}

For our purpose, we derive the Hermitian operator in two steps by the
techniques of integration within an ordered product of operators normal
ordered technique \cite{30}. Let us start by obtaining the normal ordered
form of the unitary operator related with the whole SU(1,1) interferometer.
The action of the OPA on a two-mode state is described by a two-mode
squeezing operator $\hat{S}_{2}\left( \xi \right) =\exp \left( \xi \hat{a}%
^{\dagger }\hat{b}^{\dagger }-\xi ^{\ast }\hat{a}\hat{b}\right) $ with
squeezing parameter $\xi =ge^{i\theta }$, where $g$ and $\theta $ are the
parametric gain and phase of the OPA, respectively. According to Eq. (3.66)
in Ref. \cite{30a}, the most useful factored form of the operator $%
S_{2}\left( \xi \right) $ is 
\begin{eqnarray}
\hat{S}_{2}\left( \xi \right) &=&\mathrm{sech}g\exp \left[ \hat{a}^{\dagger }%
\hat{b}^{\dagger }e^{i\theta }\tanh g\right]  \notag \\
&&\colon \exp \left[ -(\hat{a}^{\dagger }\hat{a}+\hat{b}^{\dagger }\hat{b}%
)\left( 1-\mathrm{sech}g\right) \right] \colon  \notag \\
&&\exp [-\hat{a}\hat{b}e^{-i\theta }\tanh g],  \label{1}
\end{eqnarray}%
where the notation $\colon $ $\colon $stands for the normal ordered form of
operators, which means all the Bosonic creation operators $\hat{a}^{\dag }$ (%
$\hat{b}^{\dag }$) standing on the left of annihilation operators $\hat{a}$ (%
$\hat{b}$) in a monomial of $\hat{a}$ ($\hat{b}$) and $\hat{a}^{\dag }$ ($%
\hat{b}^{\dag }$) \cite{30,31}. Within the normally ordered product of
operators, the order of the Bosonic operators can be exchanged without
affects on the result.

After the first OPA of the SU (1,1) interferometer, mode $a$ (or $b$) is
retained as a reference, while the mode $b$ (or $a$) experiences a phase
shift $\phi $. After the two modes recombine in the second OPA, the outputs
of the two modes are dependent on the phase difference $\phi $. According to
Ref. \cite{13}, the unitary transformation associated with such
interferometer can be represented by the following unitary operator 
\begin{equation}
\hat{U}\left( \xi ,\phi \right) =\hat{S}_{2}\left( -\xi \right) e^{i\phi 
\hat{a}^{\dagger }\hat{a}}\otimes \hat{I}_{b}\hat{S}_{2}\left( \xi \right) .
\label{2}
\end{equation}%
Here, the unkwon phase shift occurs only in mode $a$. For our purpose, it is
useful to express the operator $\hat{U}\left( \xi ,\phi \right) $ in the
normal ordered form. Based on the coherent state representation, the phase
shift operator $e^{i\phi \hat{a}^{\dagger }\hat{a}}$ can be expressed as 
\cite{32} 
\begin{equation}
e^{i\phi \hat{a}^{\dagger }\hat{a}}=\int \frac{d^{2}\alpha }{\pi }|\alpha
\rangle _{a}{_{a}\langle e^{-i\phi }\alpha |}.  \label{3}
\end{equation}%
Noting that the integral formula \cite{33} 
\begin{equation}
\int \frac{d^{2}z}{\pi }e^{\zeta \left\vert z\right\vert ^{2}+\xi z+\eta
z^{\ast }}=-\frac{1}{\zeta }e^{-\frac{\xi \eta }{\zeta }},  \label{4}
\end{equation}%
whose convergent condition is Re$\left( \zeta \right) <0$, by substituting
the unit operator $\hat{I}_{b}=\int d^{2}\beta |\beta \rangle _{b}{%
_{b}\langle \beta |}/\pi $ in the coherent state representation and Eq. (\ref%
{3}) into Eq. (\ref{2}), we can directly perform the integration and derive
the normal ordered form of the unitary operator $\hat{U}\left( \xi ,\phi
\right) $ (See Appendix A)%
\begin{eqnarray}
\hat{U}\left( \xi ,\phi \right) &=&\frac{1}{\cosh ^{2}g-e^{i\phi }\sinh ^{2}g%
}\exp \left( \hat{a}^{\dagger }\hat{b}^{\dagger }e^{i\theta }\tanh gA\right)
\notag \\
&&\colon \exp \left( \hat{a}^{\dagger }\hat{a}A+\hat{b}^{\dagger }\hat{b}%
B\right) \colon \exp \left( \hat{a}\hat{b}e^{-i\theta }\tanh gA\right) ,
\label{5}
\end{eqnarray}%
with%
\begin{equation}
A=\frac{\left( e^{i\phi }-1\right) \cosh ^{2}g}{\cosh ^{2}g-e^{i\phi }\sinh
^{2}g},B=\frac{\left( e^{i\phi }-1\right) \sinh ^{2}g}{\cosh ^{2}g-e^{i\phi
}\sinh ^{2}g}.  \label{6}
\end{equation}%
Naturally, when an arbitrary state passes through such SU(1,1)
interferometer, the output state can be written as%
\begin{equation}
\rho _{\text{out}}=\hat{U}\left( \xi ,\phi \right) \hat{\rho}_{\text{in}}%
\hat{U}^{\dagger }\left( \xi ,\phi \right)  \label{7}
\end{equation}%
According to Eq. (\ref{3}), when $\phi =0$, the output state $\hat{\rho}_{%
\text{out}}$ is the same as the input state $\rho _{\text{in}}$ as expected.

\subsection{Equivalent Hermitian operator of parity measurement combined
within an SU(1,1) interferometer}

Now, we turn to derive the equivalent Hermitian operator of the parity
measurement combined within a lossless SU(1,1) interferometer. In the
previous traditional operator method, the input state $\hat{\rho}_{\text{in}%
} $ passes through an optical interferometer and evolves into the output
state $\hat{\rho}_{\text{out}}$. And then, one perform a concrete
measurement of some observables $\hat{O}$ at the output state of such
devices, i.e., Tr$\left( \hat{\rho}_{\text{out}}\hat{\Pi}_{b}\right) $ with
the measurement operator $\hat{O}$. In general, one can adopt the amplitude
quadrature $\hat{X}$, photon number $\hat{N}$, and parity operator $\hat{\Pi}
$ as a measurement operator. Here, we consider the parity measurement. It is
well known that the parity measurement at one output of the interferometer
is equivalent to the expectation value of the parity operator (for example
on mode $b$, $\hat{\Pi}_{b}=\exp \left( i\pi \hat{b}^{\dag }\hat{b}\right) $%
), i.e.,%
\begin{equation}
\left\langle \hat{\Pi}\left( \phi \right) \right\rangle =\text{Tr}\left( 
\hat{\rho}_{\text{out}}\hat{\Pi}_{b}\right) .  \label{8}
\end{equation}%
Different from the previous traditional operator method, in this work we
consider the parity measurement in Heisenberg representation. Substituting
Eq. (\ref{7}) into Eq. (\ref{8}), we obtain the parity signal as%
\begin{equation}
\left\langle \hat{\Pi}\left( \phi \right) \right\rangle =\text{Tr}\left[ 
\hat{\rho}_{\text{in}}\hat{\mu}\left( \xi ,\phi \right) \right] ,  \label{a1}
\end{equation}%
where we introduce a new measurement operator $\hat{\mu}\left( \xi ,\phi
\right) $ defined by 
\begin{equation}
\hat{\mu}\left( \xi ,\phi \right) =\hat{U}^{\dagger }\left( \xi ,\phi
\right) \left[ \hat{I}_{a}\otimes \exp \left( i\pi \hat{b}^{\dag }\hat{b}%
\right) \right] \hat{U}\left( \xi ,\phi \right) .  \label{9}
\end{equation}%
Therefore, in terms of the input state, the signal of parity measurement can
be also obtained in principle. Obviously, the measurement operator $\hat{\mu}%
\left( \xi ,\phi \right) $ is a Hermitian operator, which can completely
represent the operation of the parity measurement combined with an SU(1,1)
interferometer. For our purpose, in this following work we mainly focus on
the normal ordered form of the Hermitian operator $\hat{\mu}\left( \xi ,\phi
\right) $.

Based on the coherent state representation, the parity operator can be
expressed as \cite{32} 
\begin{equation}
\left( -1\right) ^{\hat{b}^{\dag }\hat{b}}=\exp \left( i\pi \hat{b}^{\dag }%
\hat{b}\right) =\int \frac{d^{2}\beta }{\pi }|\beta \rangle _{b}{_{b}\langle
-\beta |}.  \label{10}
\end{equation}%
Similarly to the calculation of Eq. (\ref{5}), substituting the unit
operator $\hat{I}_{a}=\int d^{2}|\alpha \rangle _{a}\,{_{a}\langle \alpha |}%
/\pi $ and Eq. (\ref{9}) into Eq. (\ref{10}), we perform the integration and
finally obtain the normal ordered form of such Hermitian operator (See
Appendix B) 
\begin{eqnarray}
\hat{\mu}\left( \xi ,\phi \right) &=&\frac{1}{1+2\sin ^{2}\frac{\varphi }{2}%
\sinh ^{2}2g}\exp \left( \hat{a}^{\dagger }\hat{b}^{\dagger }M^{\ast }\right)
\notag \\
&&\colon \exp \left( -\hat{a}^{\dagger }\hat{a}C-\hat{b}^{\dagger }\hat{b}%
D\right) \colon \exp \left( \hat{a}\hat{b}M\right) ,  \label{11}
\end{eqnarray}%
where%
\begin{equation}
M=\frac{e^{-i\theta }\left( i\sin \phi -2\sin ^{2}\frac{\phi }{2}\cosh
2g\right) \sinh 2g}{1+2\sin ^{2}\frac{\phi }{2}\sinh ^{2}2g},  \label{12}
\end{equation}%
\begin{equation}
C=\frac{2\sin ^{2}\frac{\phi }{2}\sinh ^{2}2g}{1+2\sin ^{2}\frac{\phi }{2}%
\sinh ^{2}2g},D=\frac{2+2\sin ^{2}\frac{\phi }{2}\sinh ^{2}2g}{1+2\sin ^{2}%
\frac{\phi }{2}\sinh ^{2}2g},  \label{13}
\end{equation}%
with the relation $CD=\left\vert M\right\vert ^{2}$.

In this way, we obtain the normal ordered form of the Hermitian operator $%
\hat{\mu}\left( \xi ,\phi \right) $. Noting the eigenvalue equations of
annihilation operator $\hat{a}\left\vert \alpha \right\rangle =\alpha
\left\vert \alpha \right\rangle \,$\ ($\left\langle \alpha \right\vert \hat{a%
}^{\dagger }=\left\langle \alpha \right\vert \alpha ^{\ast })$, if one cast
the input state in the coherent state representation, it is convenient to
derive the signal of the parity measurement based on Eqs. (\ref{a1}) and (%
\ref{11}). For example, when a two-mode vacuum state $\rho _{\text{in}%
}=\left\vert 0\right\rangle _{a}\left\vert 0\right\rangle _{bb}\left\langle
0\right\vert _{a}\left\langle 0\right\vert $ is injected into an SU(1,1)
interferometer, the parity signal can be immediately obtained 
\begin{equation}
\left\langle \hat{\Pi}\left( \phi \right) \right\rangle =\frac{1}{1+2\sin
^{2}\frac{\phi }{2}\sinh ^{2}2g}.  \label{a2}
\end{equation}%
Noting that the phase sensitivity using the parity measurement is derived by
the error propagation theory, $\Delta \phi =\left\langle \Delta \hat{\Pi}%
_{b}\left( \phi \right) \right\rangle /\left\vert \partial \left\langle
\Delta \hat{\Pi}_{b}\left( \phi \right) \right\rangle /\partial \phi
\right\vert $, one can easily obtain the phase sensitivity with parity
measurement $1/\sqrt{2\sinh ^{2}g\left( 2\sinh ^{2}g+2\right) }$, which is
the same as the result of Yurke's scheme with intensity measurement \cite{13}%
. Further, if one considers a two-mode coherent state, $\rho _{\text{in}%
}=\left\vert \alpha \right\rangle _{a}\left\vert \beta \right\rangle
_{bb}\left\langle \beta \right\vert _{a}\left\langle \alpha \right\vert $,
as the input state of the SU(1,1) interferometer, the parity signal reads 
\begin{eqnarray}
\left\langle \hat{\Pi}\left( \phi \right) \right\rangle &=&\frac{1}{1+2\sin
^{2}\frac{\phi }{2}\sinh ^{2}2g}  \notag \\
&&\exp \left[ 2\mathrm{Re}(\alpha \beta M)-\left\vert \alpha \right\vert
^{2}C-\left\vert \beta \right\vert ^{2}D\right] .  \label{a3}
\end{eqnarray}%
Compared with that result in Ref. \cite{27}, the parity signal given by Eq. (%
\ref{a3}) is concise and illuminating expression. In the case of $\alpha
=\beta =0$, Eq. (\ref{a3}) naturally reduces to Eq. (\ref{a2}).

\section{Some Applications}

Here, we present a new method for obtaining the signal of the parity
measurement within an SU(1,1) interferometer directly in terms of input
states. In quantum optics, some Gaussian states can be express by positive $%
P $-representation, i.e., $\rho =\int d^{2}\alpha P\left( \alpha \right)
\left\vert \alpha \right\rangle \left\langle \alpha \right\vert /\pi $,
where $\left\vert \alpha \right\rangle $ is a coherent state \cite{14,34}.
On the other hand, for Gaussian or non-Gaussian states, they can be always
expressed in the coherent state representation, for example $\left\vert \psi
\right\rangle =\int d^{2}\alpha \left\vert \alpha \right\rangle \left\langle
\alpha \right\vert \left\vert \psi \right\rangle /\pi $. Based on our
method, it is relatively easy to calculate the signal of the parity
measurement in an optical interferometer. In order to show the advantages of
our method, in what follows we consider two specific states, i.e., Gaussian
states and non-Gaussian states.

\subsection{Coherent state and squeezed vacuum state}

The squeezed vacuum state (SVS) is a Gaussian state, $\left\vert
r\right\rangle _{b}=S\left( r\right) \left\vert 0\right\rangle _{b}$, where
the single-mode squeezing operator $S\left( r\right) =\mathrm{sech}%
^{1/2}r\exp \left[ \left( re^{-i\theta _{s}}\hat{b}^{2}-re^{i\theta _{s}}%
\hat{b}^{\dagger 2}\right) /2\right] $ with the squeezing parameter $r$. For
the convenience of the latter calculation, we rewrite the SVS $\left\vert
r\right\rangle _{b}$ in the basis of the coherent state as follows 
\begin{equation}
\left\vert r\right\rangle _{b}=\mathrm{sech}^{1/2}r\int \frac{d^{2}\beta }{%
\pi }e^{-\frac{1}{2}\left\vert \beta \right\vert ^{2}-\frac{\tanh r}{2}%
e^{i\theta _{s}}\beta ^{\ast 2}}\left\vert \beta \right\rangle _{b},
\label{15}
\end{equation}%
where we have used Eq. (\ref{1}) and the completeness of the coherent state$%
\int d^{2}\beta \left\vert \beta \right\rangle _{b}\left\langle \beta
\right\vert /\pi =\hat{I}_{b}$.

When we consider a coherent state and a SVS, $\hat{\rho}_{\text{in}%
}=\left\vert \alpha \right\rangle _{a}\left\langle \alpha \right\vert
\otimes \left\vert r\right\rangle _{b}\left\langle r\right\vert $, as the
input state of the SU(1,1), according to Eqs. (\ref{a1}) and (\ref{11}) we
obtain the parity signal after strait (See Appendix C) 
\begin{eqnarray}
\left\langle \hat{\Pi}\left( \phi \right) \right\rangle _{0} &=&\frac{\left(
1+2\sin ^{2}\frac{\phi }{2}\sinh ^{2}2g\right) ^{-1}}{\sqrt{\cosh
^{2}r-\left( D-1\right) ^{2}\sinh ^{2}r}}  \notag \\
&&\times \exp \left[ -\frac{\left\vert \alpha \right\vert ^{2}(C+\left\vert
M\right\vert ^{2}\sinh ^{2}r)}{\cosh ^{2}r-\left( D-1\right) ^{2}\sinh ^{2}r}%
\right.  \notag \\
&&\left. -\frac{\mathrm{Re}\left[ \alpha ^{2}M^{2}e^{i\theta _{s}}\right]
\sinh 2r}{2\left( \cosh ^{2}r-\left( D-1\right) ^{2}\sinh ^{2}r\right) }%
\right] .  \label{16}
\end{eqnarray}%
Compared with that result in Ref. \cite{27}, the parity signal given by Eq. (%
\ref{16}) remains to be concise and illuminating expression.

\subsection{A thermal state and squeezed vacuum state}

It is known that the $P$-representation of density operator of a thermal
state is%
\begin{equation}
\rho _{\text{th}}=\frac{1}{\bar{n}_{\text{th}}}\int \frac{d^{2}\alpha }{\pi }%
\exp \left( -\frac{1}{\bar{n}_{\text{th}}}\left\vert \alpha \right\vert
^{2}\right) \left\vert \alpha \right\rangle _{a}\left\langle \alpha
\right\vert .  \label{17}
\end{equation}%
Therefore, when further considering a thermal state and a SVS, $\hat{\rho}_{%
\text{in}}=\rho _{\text{th}}\otimes \left\vert r\right\rangle
_{b}\left\langle r\right\vert $, as the input state of the SU(1,1),
according to Eqs. (\ref{a1}) and (\ref{16}) we can easily obtain the signal
of parity measurement within SU(1,1)%
\begin{eqnarray}
&&\left\langle \hat{\Pi}\left( \phi \right) \right\rangle  \notag \\
&=&\frac{1}{\bar{n}_{\text{th}}}\int \frac{d^{2}\alpha }{\pi }\exp \left( -%
\frac{1}{\bar{n}_{\text{th}}}\left\vert \alpha \right\vert ^{2}\right)
\left\langle \hat{\Pi}\left( \phi \right) \right\rangle _{0}  \notag \\
&=&\frac{\left( 1+2\sin ^{2}\frac{\phi }{2}\sinh ^{2}2g\right) ^{-1}}{\sqrt{%
\cosh ^{2}r-\left( D-1\right) ^{2}\sinh ^{2}r}}\times  \label{18} \\
&&\frac{1}{\sqrt{\left[ 1+\frac{C\mathrm{sech}^{2}r+\left\vert M\right\vert
^{2}\tanh ^{2}r}{1-\left( D-1\right) ^{2}\tanh ^{2}r}\bar{n}_{\text{th}}%
\right] ^{2}-\frac{\bar{n}_{\text{th}}^{2}\left\vert M\right\vert ^{4}\tanh
^{2}r}{\left[ 1-\left( D-1\right) ^{2}\tanh ^{2}r\right] ^{2}}}}.  \notag
\end{eqnarray}%
Compared with that result in Ref. \cite{28}, the parity signal given by Eq. (%
\ref{18}) is more concise and illuminating. Obviously, one can see from the
above two cases that the new operator method expressed by Eqs. (\ref{a1})
and (\ref{11}) is more convenient to obtain the parity signal of an SU(1,1)
interferometer than the phase space method used in Ref. \cite{27,28}.

\subsection{A Fock state}

Finally, we consider a typical kind of a non-Gaussian state, that is a Fock
state. Mathematically, the Fock state $\left\vert n\right\rangle _{b}=\left( 
\hat{b}^{\dagger n}/\sqrt{n!}\right) \left\vert 0\right\rangle _{b}$ can be
expressed by $\left\vert n\right\rangle _{b}=\left( \partial ^{n}/\sqrt{n!}%
\partial x^{n}\right) \exp \left[ x\hat{b}^{\dagger }\right] \left\vert
0\right\rangle _{b}|_{x=0}$ in quantum mechanics. For the sake of
convenience, then we further rewrite the Fock state in the coherent
representation, i.e., 
\begin{equation}
\left\vert n\right\rangle _{b}=\frac{\partial ^{n}}{\sqrt{n!}\partial x^{n}}%
\int \frac{d^{2}\beta }{\pi }\exp \left( -\frac{1}{2}\left\vert \beta
\right\vert ^{2}+x\beta ^{\ast }\right) \left\vert \beta \right\rangle _{b}.
\label{19}
\end{equation}%
If a vacuum state and a Fock state $\rho _{\text{in}}=\left\vert
0\right\rangle _{a}\left\langle 0\right\vert \otimes \left\vert
n\right\rangle _{b}\left\langle n\right\vert $ are considered as the input
state of an SU(1,1) interferometer, then the parity signal can be also
immediately obtained,

\begin{equation}
\left\langle \hat{\Pi}\left( \phi \right) \right\rangle =\left( 1+2\sin ^{2}%
\frac{\phi }{2}\sinh ^{2}2g\right) ^{-(n+1)},  \label{20}
\end{equation}%
which is a new result. Based on Eq. (\ref{20}), we can further investigate
the phase sensitivity of the SU(1,1) interferometer with the Fock state as
an input state. In the case of $n=1$, Eq. (\ref{20}) reduces to that result
in Ref. \cite{25}. In addition, based on Eqs. (\ref{a1}) and (\ref{11}), the
parity signal can be also obtained when a coherent or thermal state and a
Fock state as the input state of an SU(1,1) interferometer. And then, the
phase sensitivity can be investigated by the error propagation theorem.
Here, we don't discuss these in detail. One can see again that the new
operator method proposed in this work may be an effective way in quantum
metrology.

\section{Conclusions}

In summary, we have derived a Hermitian operator which is equal to the
operation of the whole SU(1,1) interferometer combined with the parity
measurement. Different from the previous traditional operator method or the
phase space method, we propose a new operator method in the Heisenberg
representation by which one can obtain the signal of the parity measurement
within an SU(1,1) interferometer directly based on the input states. By this
new method, it is relatively simpler to calculate the signal of the parity
measurement in the SU(1,1) interferometer with Gaussian or non-Gaussian
states. Our work may bring convenience to the study on quantum metrology,
particularly the phase estimation based on SU(1,1) interferometers.

\section*{Acknowledgments}

This work is supported by the National Natural Science Foundation of China
(Grant No. 11404040), and the Natural Science Foundation of Jiangsu Province
of China (Grant No. BK20140253).

\section*{Appendix A}

Substituting the unit operator $\hat{I}_{b}=\int d^{2}|\beta \rangle _{b}{%
_{b}\langle \beta |}/\pi $ and Eq. (\ref{3}) into Eq. (\ref{2}), we have%
\begin{equation}
\hat{U}\left( \xi ,\phi \right) =\int \frac{d^{2}\alpha _{1}d^{2}\alpha _{2}%
}{\pi ^{2}}S_{2}\left( -\xi \right) |\alpha _{1}\rangle _{a}\,\left\vert
\alpha _{2}\right\rangle _{b}\,{_{b}}\left\langle \alpha _{2}\right\vert {%
_{a}\langle e^{-i\phi }\alpha _{1}|}S_{2}\left( \xi \right)  \tag{A1}
\end{equation}%
Noting that the normal ordering form of the two-mode squeezing operator Eq. (%
\ref{1}), and the eigenvalue equations $\hat{a}\left\vert \alpha
\right\rangle _{a}=\alpha \left\vert \alpha \right\rangle _{a}$ as well as $%
\hat{b}|\beta \rangle _{b}=\beta |\beta \rangle _{b}$, then we obtain%
\begin{align}
\hat{U}\left( \xi ,\phi \right) & =\mathrm{sech}^{2}g\int \frac{d^{2}\alpha
_{1}d^{2}\alpha _{2}}{\pi ^{2}}\colon \exp \left[ -\left\vert \alpha
_{1}\right\vert ^{2}-\left\vert \alpha _{2}\right\vert ^{2}\right.  \notag \\
& \left. +(\hat{a}^{\dagger }\alpha _{1}+\hat{b}^{\dagger }\alpha _{2})%
\mathrm{sech}g+\alpha _{1}\alpha _{2}e^{-i\theta }\tanh g\right.  \notag \\
& \left. (e^{i\phi }\alpha _{1}^{\ast }\hat{a}+\alpha _{2}^{\ast }\hat{b})%
\mathrm{sech}g+e^{i\phi }\alpha _{1}^{\ast }\alpha _{2}^{\ast }e^{i\theta
}\tanh g\right.  \notag \\
& \left. -\hat{a}^{\dagger }\hat{b}^{\dagger }e^{i\theta }\tanh g-\hat{a}%
\hat{b}e^{-i\theta }\tanh g-\hat{a}^{\dagger }\hat{a}-\hat{b}^{\dagger }\hat{%
b}\right] \colon ,  \tag{A2}
\end{align}%
where we have used the operator identity $\left\vert 0\right\rangle
_{a}\left\vert 0\right\rangle _{bb}\left\langle 0\right\vert
_{a}\left\langle 0\right\vert =\colon \exp \left[ -\hat{a}^{\dagger }\hat{a}-%
\hat{b}^{\dagger }\hat{b}\right] \colon $. According to those properties of
normally ordered product of operators, when the operator function $F\left( 
\hat{a},\hat{a}^{\dag },\hat{b},\hat{b}^{\dag }\right) $ is converted to the
normal ordering, one can treat operators $\hat{a}$ ($\hat{b}$) and $\hat{a}%
^{\dag }$ ($\hat{b}^{\dag }$) in Eq. (A2) as the $c$-number parameters and
carry out the integration safely \cite{30}. Applying the integration formula
Eq. (\ref{4}), we can directly perform the integration of Eq. (A2) over the
whole of the complex plane and then obtain 
\begin{align}
\hat{U}\left( \xi ,\phi \right) & =\frac{1}{\cosh ^{2}g-e^{i\phi }\sinh ^{2}g%
}\colon \exp \left( \hat{a}^{\dagger }\hat{b}^{\dagger }e^{i\theta }\tanh
gA\right.  \notag \\
& \left. +\hat{a}^{\dagger }\hat{a}A+\hat{b}^{\dagger }\hat{b}B+\hat{a}\hat{b%
}e^{-i\theta }\tanh gA\right) \colon ,  \tag{A3}
\end{align}%
According to those properties of normally ordered product of operators, we
can further convert Eq. (A3) to Eq. (\ref{5}).

\section*{Appendix B}

Now, we turn to derive the corresponding Hermitian operator for parity
measurement within an SU(1,1) interferometer. Substituting the unit operator 
$\hat{I}_{b}=\int d^{2}\alpha _{1}\left\vert \alpha _{1}\right\rangle
_{a}\left\langle \alpha _{1}\right\vert /\pi $ and Eq. (\ref{10}) into Eq. (%
\ref{9}), we have%
\begin{equation}
\hat{\mu}\left( \xi ,\phi \right) =\int \frac{d^{2}\alpha d^{2}\beta }{\pi
^{2}}\hat{U}^{\dagger }\left( \xi ,\phi \right) |\alpha \rangle
_{a}\,\left\vert \beta \right\rangle _{b}\,{_{b}}\left\langle {-\beta }%
\right\vert {_{a}\langle \alpha |}\hat{U}\left( \xi ,\phi \right)  \tag{B1}
\end{equation}%
Noting that the normal ordering form of the unitary operator $\hat{U}\left(
\xi ,\phi \right) $, similarly to derive Eq. (\ref{5}), we obtain

\begin{align}
\hat{\mu}\left( \xi ,\phi \right) & =\frac{1}{1+\sin ^{2}\frac{\phi }{2}%
\sinh ^{2}2g}\int \frac{d^{2}\alpha d^{2}\beta }{\pi ^{2}}\colon \exp \left[
-\left\vert \alpha \right\vert ^{2}\right.  \notag \\
& \left. -\left\vert \beta \right\vert ^{2}+\alpha \hat{a}^{\dagger }+\alpha
^{\ast }\hat{a}+\alpha \hat{a}^{\dagger }A^{\ast }+{\alpha }^{\ast }\hat{a}%
A+\beta \hat{b}^{\dagger }\right.  \notag \\
& \left. -\beta ^{\ast }\hat{b}+\beta \hat{b}^{\dagger }B^{\ast }-\beta
^{\ast }\hat{b}B+\alpha \beta e^{-i\theta }A^{\ast }\tanh g\right.  \notag \\
& \left. -{\alpha }^{\ast }\beta ^{\ast }e^{i\theta }A\tanh g+\hat{a}\hat{b}%
e^{-i\theta }A\tanh g\right.  \notag \\
& \left. +\hat{a}^{\dagger }\hat{b}^{\dagger }e^{i\theta }A^{\ast }\tanh g-%
\hat{a}^{\dagger }\hat{a}-\hat{b}^{\dagger }\hat{b}\right] \colon .  \tag{B2}
\end{align}%
Then, by the the integration formula Eq. (\ref{4}), we can directly perform
the integration of Eq. (B2) and finally obtain Eq. (\ref{11}).

\section*{Appendix C}

When we consider a coherent state and an SVS, $\hat{\rho}_{\text{in}%
}=\left\vert \alpha \right\rangle _{a}\left\langle \alpha \right\vert
\otimes \left\vert r\right\rangle _{b}\left\langle r\right\vert $, as the
input state of the SU(1,1), according to Eqs. (\ref{a1}) and (\ref{11}) we
obtain the signal of parity measurement within an SU(1,1) interferometer,%
\begin{align}
\left\langle \hat{\Pi}_{b}\left( \phi \right) \right\rangle & =\frac{\mathrm{%
sech}r}{1+2\sin ^{2}\frac{\phi }{2}\sinh ^{2}2g}\int \frac{d^{2}\beta
_{1}d^{2}\beta _{2}}{\pi ^{2}}  \notag \\
& e^{-\frac{1}{2}\left\vert \beta _{2}\right\vert ^{2}-\frac{1}{2}\left\vert
\beta _{1}\right\vert ^{2}-\frac{\tanh r}{2}e^{i\theta _{s}}\beta _{1}^{\ast
2}-\frac{\tanh r}{2}e^{-i\theta _{s}}\beta _{2}^{2}}  \notag \\
& \left. _{b}\left\langle \beta _{2}\right\vert \right. _{a}\left\langle
\alpha \right\vert e^{\hat{a}^{\dagger }\hat{b}^{\dagger }M^{\ast }}\colon
e^{-\hat{a}^{\dagger }\hat{a}C-\hat{b}^{\dagger }\hat{b}D}\colon e^{\hat{a}%
\hat{b}M}\left\vert \alpha \right\rangle _{a}\left\vert \beta
_{1}\right\rangle _{b}.  \tag{C1}
\end{align}%
Then noting that the eigenvalue equations $\hat{a}\left\vert \alpha
\right\rangle _{a}=\alpha \left\vert \alpha \right\rangle _{a}$ and $\hat{b}%
|\beta \rangle _{b}=\beta |\beta \rangle _{b}$, as well as the
non-orthogonality relation of the coherent state $\left\langle \beta
_{2}\right\vert \left. \beta _{1}\right\rangle =\exp \left( -\frac{1}{2}%
\left\vert \beta _{1}\right\vert ^{2}-\frac{1}{2}\left\vert \beta
_{2}\right\vert ^{2}+\beta _{2}^{\ast }\beta _{1}\right) $, we have%
\begin{align}
\left\langle \hat{\Pi}_{b}\left( \phi \right) \right\rangle & =\frac{\mathrm{%
sech}r}{1+2\sin ^{2}\frac{\phi }{2}\sinh ^{2}2g}\int \frac{d^{2}\beta
_{1}d^{2}\beta _{2}}{\pi ^{2}}  \notag \\
& \exp \left[ -\left\vert \alpha \right\vert ^{2}C-\left\vert \beta
_{1}\right\vert ^{2}-\left\vert \beta _{2}\right\vert ^{2}\right.  \notag \\
& \left. +\alpha ^{\ast }\beta _{2}^{\ast }M^{\ast }+\alpha \beta
_{1}M-\beta _{2}^{\ast }\beta _{1}\left( D-1\right) \right.  \notag \\
& \left. -\frac{\tanh r}{2}e^{i\theta _{s}}\beta _{1}^{\ast 2}-\frac{\tanh r%
}{2}e^{-i\theta _{s}}\beta _{2}^{2}\right]  \tag{C2}
\end{align}%
By the following integral formula \cite{33} 
\begin{equation}
\int \frac{d^{2}z}{\pi }e^{\zeta \left\vert z\right\vert ^{2}+\xi z+\eta
z^{\ast }+fz^{2}+gz^{\ast 2}}=\frac{1}{\sqrt{\zeta ^{2}-4fg}}e^{\frac{-\zeta
\xi \eta +\xi ^{2}g+\eta ^{2}f}{\zeta ^{2}-4fg}},  \tag{C3}
\end{equation}%
whose convergent condition is Re$\left( \zeta \pm f\pm g\right) <0,\ $Re$%
\left( \frac{\zeta ^{2}-4fg}{\zeta \pm f\pm g}\right) <0$, we can directly
perform the integration of Eq. (C2) and finally obtain Eq. (\ref{16}). In
the last step of deriving Eq. (\ref{16}), we have used the relation $%
CD=\left\vert M\right\vert ^{2}$.


\end{document}